\documentclass[useAMS,usenatbib]{mn2e}
\usepackage{graphicx}
\usepackage{subfigure}

\newcommand{\kms}{km~s\ensuremath{^{-1}}}

\newcommand{\ms}{m~s\ensuremath{^{-1}}}

\def\ms{m$\,$s$^{-1}$}

\title[WASP-12]{Orbital eccentricity of WASP-12 and WASP-14 from new radial-velocity monitoring with SOPHIE
\thanks{Based on observations 
   made at the 1.93-m telescopes at Observatoire de Haute-Provence (CNRS), 
   France with the SOPHIE spectrograph.}
}

\begin{document}
\bibliographystyle{mn2e}

%
%

\author[N. Husnoo et al.]{Nawal Husnoo$^1$, Fr\'ed\'eric Pont$^1$, Guillaume H\'ebrard$^2$, Elaine Simpson$^3$, \newauthor Tsevi Mazeh$^4$, Fran\c cois Bouchy$^{2,5}$, Claire Moutou$^{6}$, Luc Arnold$^5$,\newauthor Isabelle Boisse$^2$, Rodrigo Diaz$^2$, Anne Eggenberger$^7$ and Avi Shporer$^{8}$\\
$^1$ School of Physics, University of Exeter, Exeter, EX4 4QL, UK \\
$^2$ Institut d'Astrophysique de Paris, UMR7095 CNRS, Universit\'e Pierre \& Marie Curie, 98bis boulevard Arago, 75014 Paris, France\\
$^3$ Astrophysics Research Centre, School of Mathematics \& Physics, Queens University, University Road, Belfast BT7 1NN, UK\\
$^4$ School of Physics and Astronomy, Tel Aviv University, Tel Aviv 69978, Israel\\
$^5$ Observatoire de Haute-Provence, CNRS/OAMP, 04870 Saint-Michel-l'Observatoire, France\\
$^6$ Laboratoire d'Astrophysique de Marseille, Universit\'e de Provence, CNRS(UMR 6110), 38 rue Fr\'ed\'eric Joliot Curie, 13388 \&\\ \ \ \  Marseille cedex 13,France\\
$^7$ Laboratoire d'Astrophysique de Grenoble, Universit\'e Joseph-Fourier, CNRS (UMR5571), BP53, 38041 Grenoble cedex 9, France\\
$^{8}$ Las Cumbres Observatory Global Telescope network, 6740 Cortona Drive, suite 102, Goleta, CA 93117,
USA\\
}




\maketitle

\label{firstpage}

\begin{abstract}

As part of the long-term radial velocity monitoring of known transiting planets -- designed to measure orbital eccentricities, spin-orbit alignments and further planetary companions -- we have acquired radial velocity data for the two transiting systems WASP-12 and WASP-14, each harbouring gas giants on close orbits (orbital period of 1.09 and 2.24 days respectively). In both cases, the initial orbital solution suggested a significant orbital eccentricity, 0.049 $\pm$ 0.015 for WASP-12 and 0.091 $\pm$ 0.003 for WASP-14. Since then, measurements of the occultation of WASP-12 in the infrared  have indicated that one projection of the eccentricity ($e \cos \omega$) was close to zero, casting doubt on the eccentricity from the initial radial velocity orbit. Our measurements confirm that the initial eccentricity detection could be spurious, and show that the radial velocity data is compatible with a circular orbit. A MCMC analysis taking into account the presence of correlated systematic noise in both the radial velocity and photometric data gives $e=0.017^{+0.015}_{-0.011}$. By contrast, we confirm the orbital eccentricity of WASP-14, and refine its value to $e=0.088\pm0.003$. WASP-14 is thus the closest presently known planet with a confirmed eccentric orbit.

\end{abstract}



\section{Introduction}

Transiting planets are an important source of information on the formation, structure and evolution of extra-solar planets. We are monitoring known transiting planetary systems in radial velocity with the SOPHIE spectrograph in the Northern hemisphere and HARPS spectrograph in the Southern hemisphere to refine our knowledge of the dynamics of these systems, notably the orbital eccentricity, spin-orbit angle and presence of additional companions  \citep[e.g.][ESO Prog. 0812.C-0312]{Loeillet2008,Hebrard2008}.

In this paper, we analyse our new SOPHIE radial-velocity data for two transiting planetary systems, WASP-12 and WASP-14. Both are characterized by close-in but apparently eccentric orbits, and therefore represent potentially important systems to constrain the migration, tidal and thermal evolution of gas giant planets. We combine our radial-velocity data with previously published data and a realistic treatment of correlated noise to calculate updated constraints on the orbital eccentricities. 


The companion of the 11.7th-magnitude star WASP-12 is a particularly interesting example \citep[][hereafter H09]{Hebb2009}. It orbits extremely close to its host star, even by the standards of the so-called ``hot Jupiters'', with a period of 1.09 days, corresponding to an orbital distance only 3 times the radius of its host star. Moreover, WASP-12b has an inflated radius, $R\simeq1.8 {\rm  \, R_J}$, one of the most extreme examples of anomalous radii for hot Jupiters. As a result, the planet fills about half of its Roche lobe \citep{Li2010}.

With such a short orbital distance and large size, a gas giant planet is expected to undergo complete orbital synchronisation and circularisation on a short timescale, much shorter than the age of a typical field main-sequence cool star. Indeed, most planets orbiting closer than 0.05 AU are observed to have circular orbits. However, H09 determined a value of $e=0.049 \pm 0.015$ for the orbital eccentricity of WASP-12, a significant departure from circularity. This would make the planet by far the subject of the strongest tidal dissipation in any known planetary system. The measured eccentricity is based on fitting a Keplerian orbital motion on the radial velocity measurements collected by H09 with the SOPHIE spectrometer \citep{Perruchot2008} together with transit photometry. 

\cite{Li2010} studied the case of WASP-12 with that value of eccentricity, and found a large implied mass loss and dissipation of tidal energy in the planet.

In a transiting system, the time lag between the transit and the occultation has a strong dependence on the projected orbital eccentricity ($e \cos \omega$). Therefore, if the occultation can be detected with sufficient significance, this provides a stringent test of the eccentricity. \citet[hereafter L09]{Morales2009} have measured the occultation of WASP-12b from the ground with SPICam on the ARC telescope at Apache Point Observatory in the $z'$ band. Their best-fit result indicated a occultation with a significant time lag compared to the epoch expected for a circular orbit, with a similar level of significance to H09. 
Nevertheless, the presence of residual correlated noise is apparent in the L09 data (see Fig.~\ref{morales-transit}), as expected for ground-based photometry at such a high accuracy - the depth of the occultation is only about 0.08$\pm$0.02 \%.  

As a result, the issue remained inconclusive until a space-based measurement of the occultation with the Spitzer Space Telescope \citep[hereafter C10]{Campo2010} unambiguously showed that the timing of the occultation was precisely that expected for a circular orbit. This result suggested that the L09 time lag was probably due to instrumental systematics, and that the orbit of WASP-12 was probably circular, since a fine-tuned alignment would be required to reconcile the Spitzer result with the H09 value of the eccentricity. 

It is interesting to note that there is an inherent bias in eccentricity measurements from radial velocities, because a Keplerian orbit cannot get more circular than $e=0$. Any noise applied to a circular orbit will result in an eccentric best-fit orbit. Underestimating the noise will lead to spurious detections of small eccentricities. This was already recognized in the context of stellar binaries by \cite{Lucy1971}. These authors showed that spurious eccentricity detections tended to dominate for $e< 0.1$ for a typical precision at that time and stellar binary amplitudes. Four decades later, both companion masses and RV accuracies having changed by about three orders of magnitudes, and the same issue resurfaces for exoplanets.


WASP-14 is, after WASP-12, the known transiting planet having a reported non-circular orbit \citep[$e=0.091 \pm 0.003$]{Joshi2009} with the second-shortest period (P=2.2 days). This makes it another test-case for tidal evolution of close-in gas giants. If its orbital eccentricity is indeed near 0.1, then this non-zero but relatively low value -- in the context of the distribution of giant exoplanet eccentricities -- makes it likely that this planet has undergone some degree of orbital evolution, and is still subject to strong tidal forces at present. Therefore its presence may be useful to constrain the tidal synchronisation timescale. It is also an important object when studying the issue of the anomalous radius of hot Jupiters because of its inflated size, with $R_p=1.28 {\rm  \, R_J}$. WASP-14 occupies a distinctive position in the relevant parameter space: irradiation, orbital distance, eccentricity and size.


\section{Observations}

We obtained 29 radial-velocity measurements for WASP-12 (16 during a single night, and 13 at various values of orbital phase) and 11 for WASP-14, using the SOPHIE spectrograph installed on the 1.93-m telescope at OHP (France). The observations were gathered between 17 January 2009 and 27 March 2010. The 16 in-transit measurements for WASP-12 were obtained with the objective of constraining the spin-orbit angle via the Rossiter-McLaughlin effect.

SOPHIE is a spectrograph optimized for precise radial-velocity measurements and has participated in the detection of numerous transiting exoplanets in the northern hemisphere, notably from the WASP and CoRoT transit searches. It reaches a stability of a few \ms\ for bright targets. WASP-12 and WASP-14, however, are near the faint end of the capacity of the 1.93-m telescope, and were measured in the ``High Efficiency'' mode of SOPHIE \citep[See][]{Perruchot2008,Bouchy2009}. This mode has a higher throughput than the standard mode, the ``High Resolution'' mode, thus allowing fainter targets to be measured, but is less optimized for radial velocity. When considering the ensemble of data for known transiting planets obtained with SOPHIE, we have found evidence for large excursions of the velocity zero-point with time (to the level of several dozen \ms\ in some cases). As part of the constant improvement of the SOPHIE reduction pipeline, this effect is monitored and corrected for as far as possible, but the presence of relatively large instrumental systematics in the High Efficiency data is a possibility, especially with older data collected before we became aware of the issue. 
This must be remembered when performing an orbital analysis based on data from the High Efficiency mode.

%
%

\section{Analysis and results}

The orbital parameters of the two transiting planets were calculated from the radial velocity data, (together with published photometry data for the transit and occultation in the case of WASP-12), with a Marko Chain Monte Carlo (MCMC) method. The main advantage of the MCMC method is that it allows a seamless combination of radial-velocity data with light-curve data both for the transit and occultation, as well as information on the parent star. The use of MCMC in this context is described by, among others, \citet{holman2006}. Our implementation is described in \cite{Pont2009b}. We use a Bayesian treatment of the {\em a priori} constraints from stellar evolution models with the method of \citet{Pont2004}. We model the radial velocity using a Keplerian orbit, the Rossiter-McLaughlin effect on the radial velocity using \citet{Gimenez2006} and the transit and eclipse light curves using \citet{Mandel2002}.

MCMC methods are powerful, but they also tend to obscure the relation between the errors on the measured data and the calculated uncertainties on the final system parameters. Neglecting non-random sources of noise in the data (such as instrumental systematics and stellar variability) can lead to an underestimation of the uncertainties in the final system parameters by a large factor \citep{Pont2006},and, in the case of orbital eccentricities, to a systematic bias \citep{Lucy1971}. We account for the presence of correlated noise in both the photometric and radial velocity data by modifying the merit function used in the MCMC to include the possible presence of non-random noise.




\subsection{The orbital eccentricity of {\small WASP}-12}



\begin{figure*}
\resizebox{8cm}{!}{\includegraphics{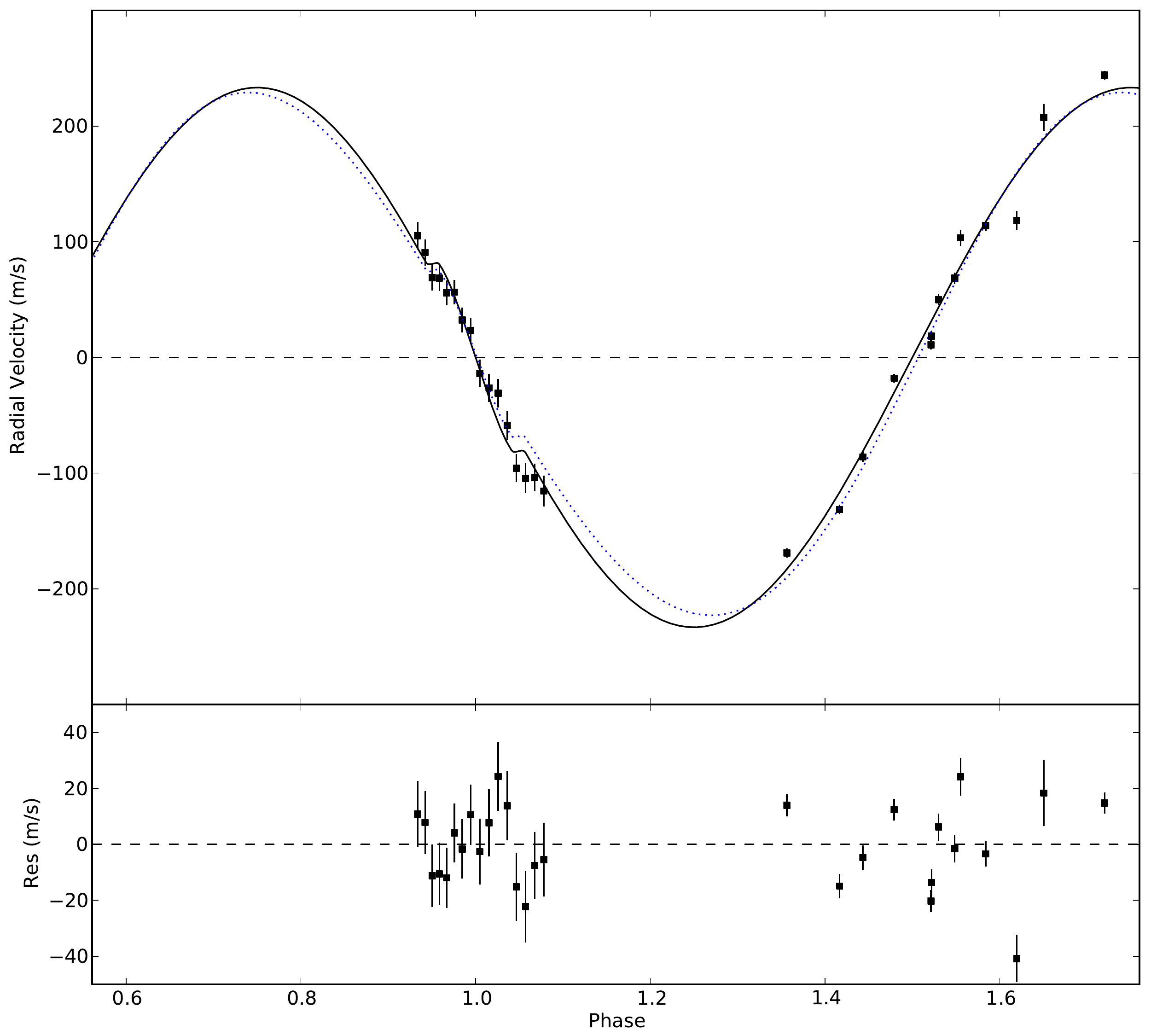}}\resizebox{8cm}{!}{\includegraphics{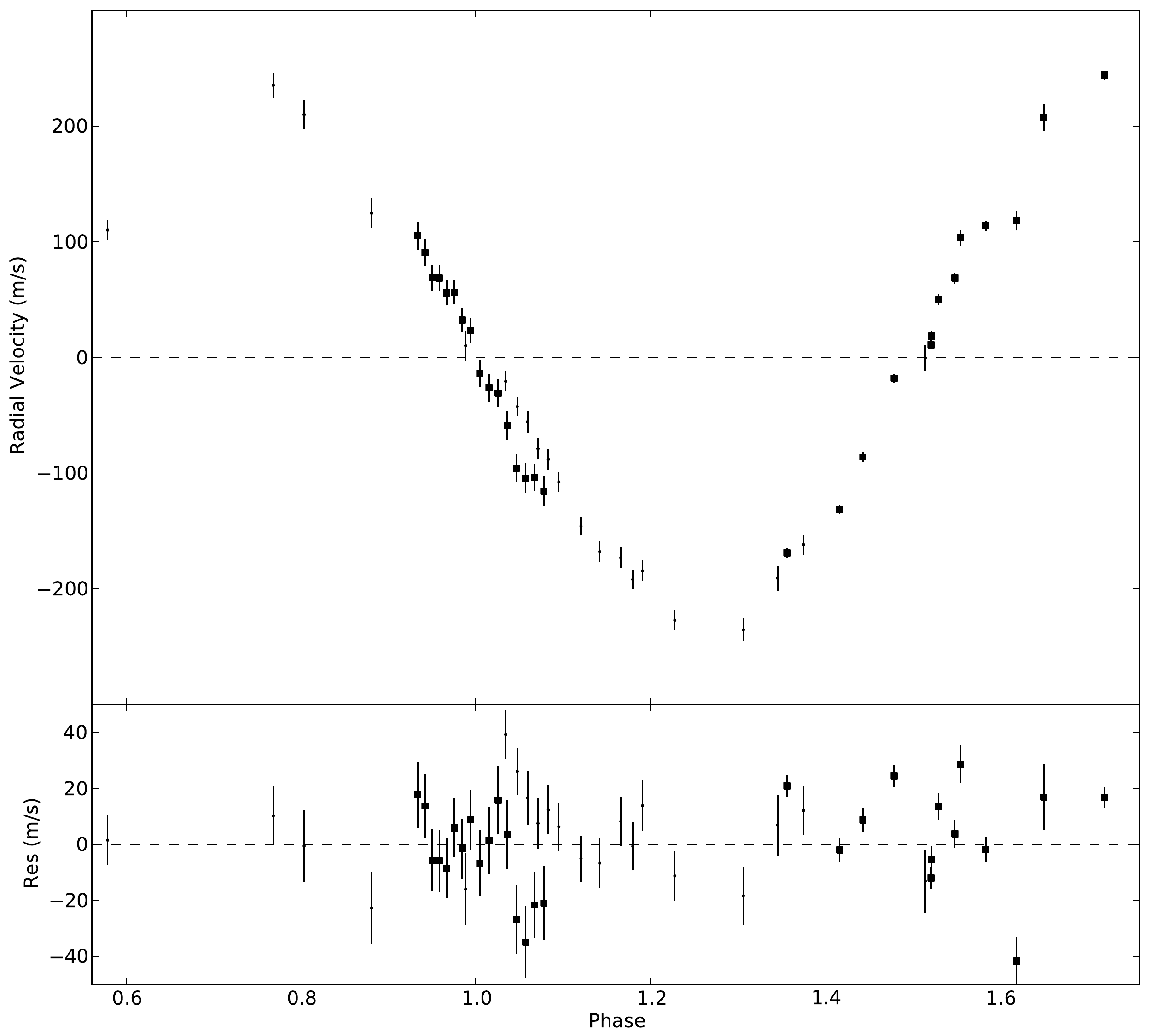}}
\caption{{\bf Left--}Plot showing our new SOPHIE radial velocity data for WASP-12, overplotted with a circular orbit (solid line) and an orbit with the eccentricity in H09. The residuals relative to the circular orbit are shown in the bottom panel.\newline
{\bf Right--}Plot showing all the radial velocity data for WASP-12, with our new SOPHIE data plotted with square symbols and the H09 data plotted with dots. Residuals for an orbit with the H09 value of eccentricity, $e=0.049$, are shown in the bottom panel.}
\label{campo-phase}
\end{figure*}

Figure~\ref{campo-phase} displays our radial-velocity data for WASP-12, together with a circular orbital solution (solid line, residuals shown in the lower panel) and the eccentric-orbit fit of H09 (dotted). Our radial velocity data is shown together with the H09 data in the right panel. As the Figure shows clearly, the radial-velocity signal cannot be adequately modelled by a periodic orbital signal affected by random noise. The presence of a correlated non-periodic component is especially obvious during the in-transit sequence. This could  be due to instrumental noise, stellar variability or an unaccounted planetary companion in the system, all of which would behave in the same way as far as the orbital fit is concerned. Given our experience with SOPHIE in High Efficiency mode, we consider the first cause as likely.


We use the transit lightcurve from H09 and the eclipse lightcurve from  L09 with our own radial velocity measurements to fit a Keplerian orbit, and we account for the effect of red noise by modifying the uncertainties on the data using
\begin{math}
\sigma ' = (\sigma^2 + N\sigma^2_{r})^{1/2}
\end{math},
where $\sigma$ is the random error, $\sigma'$ the modified error used to compute the merit function for the MCMC, $\sigma_r$ the red-noise factor, and $N$ the number of data points over a typical correlation timespan. The reader is referred to \citet{Pont2006} and Winn (2010 {\emph{in prep}}) for more details on this approach.
For the H09 photometric transit, we estimate $\sigma_r=0.0005$ and  $N=20$. 
Similarly, for the L09 data, we set $\sigma_r=0.0002$ and $N=9$. 
For our own radial velocity data, we estimate the red noise parameter to be $8$ \ms, and $N=16$ during the transit and $N=1$ outside the transit (these points are taken in different nights so that the errors are not expected to be correlated).

We vary the period $P$, mid-transit time $T_{\textrm{tr}}$, system velocity $v_0$, eccentricity components $e\cos\omega$ and $e\sin\omega$, semi-amplitude $K$, impact parameter $b$, scaled radius $R_p/R_s$, the mass $M_{\textrm{s}}$, radius $R_{\textrm{s}}$ and $T_{\textrm{eff}}$ of the star. The scaled semi-major axis $a/R_s$ is calculated from the assumption of a Keplerian orbit using the period $P$.


We set the quadratic limb darkening parameters for the H09 transit to $u_a=0.1274$  and $u_b=0.3735$ according to \cite{Claret2004}.

Our results for the orbital parameters of the WASP-12 system are shown in Table~\ref{hebb_morales_exeter}. Table~\ref{hebb_morales_exeter} shows the results of our new SOPHIE RV data used in combination with the H09 transit lightcurve, and the L09 and C10 eclipse lightcurves, compared to the H09 parameters.

\begin{table*}
\centering
\begin{tabular}{l c  c } 
\hline
Parameter										& H09							&	New SOPHIE RV and photometry	\\ \hline
Centre-of-mass velocity $V_0$ [\kms] 	& 19.085$\pm$0.002				& 19.061$\pm$0.007		  	  	\\
Orbital period $P$ [days]								& 1.091423$\pm$0.000003		& 1.09139$\pm$0.00001		\\
Orbital eccentricity $e$ 								& 0.049$\pm$0.015		& ${0.017_{-0.010}^{+0.015}}^{*}$ 			\\
Argument of periastron $\omega$ [$^o$]				& $-$74$^{+13}_{-10}$				& 0 (unconstrained)			\\ 
Velocity semi-amplitude K [m$\,$s$^{-1}$]				& 226$\pm$4					& 233$\pm$6			\\
Epoch of transit T$_{tr}$ [BJD]			& 2454508.9761$\pm$0.0003		& 2454508.9761$\pm$0.0006	\\ 
\hline
\end{tabular}
\caption{System parameters for WASP-12. Left: H09. Right: Our SOPHIE radial velocity data, H09 transit photometry data and L09 and C10  eclipse photometry data.\newline
$^*$We adopt a circular orbit for working out the rest of the system parameters. \newline
}
\label{hebb_morales_exeter}
\end{table*}

Although our data covers a complete spectroscopic transit, and thus potentially constrains the projected spin-orbit angle of the WASP-12 system through the Rossiter-McLaughlin effect, the constraint is weak once the possible presence of non-random noise is taken into account.  The distribution of the spin-orbit angle from our MCMC spans a wide interval extending from a prograde orbit to a projected spin-orbit angle larger than $90$ degrees. The data marginally favours a prograde rather than retrograde orbit.

The best solution with our radial velocity data and the constrains from the transit and eclipse light curves gives $e=0.017_{-0.010}^{+0.015}$. This is a marginal detection, so we  set $e=0$ to work out the rest of the system parameters.
The H09 radial-velocity data forces the solution towards a higher eccentricity, significant at the $\sim$3$\sigma$ level, but Figure~\ref{campo-phase} strongly suggests that this is probably an artefact due to an excursion of the zero-point between different nights.

\begin{figure*}
\resizebox{8cm}{!}{\includegraphics{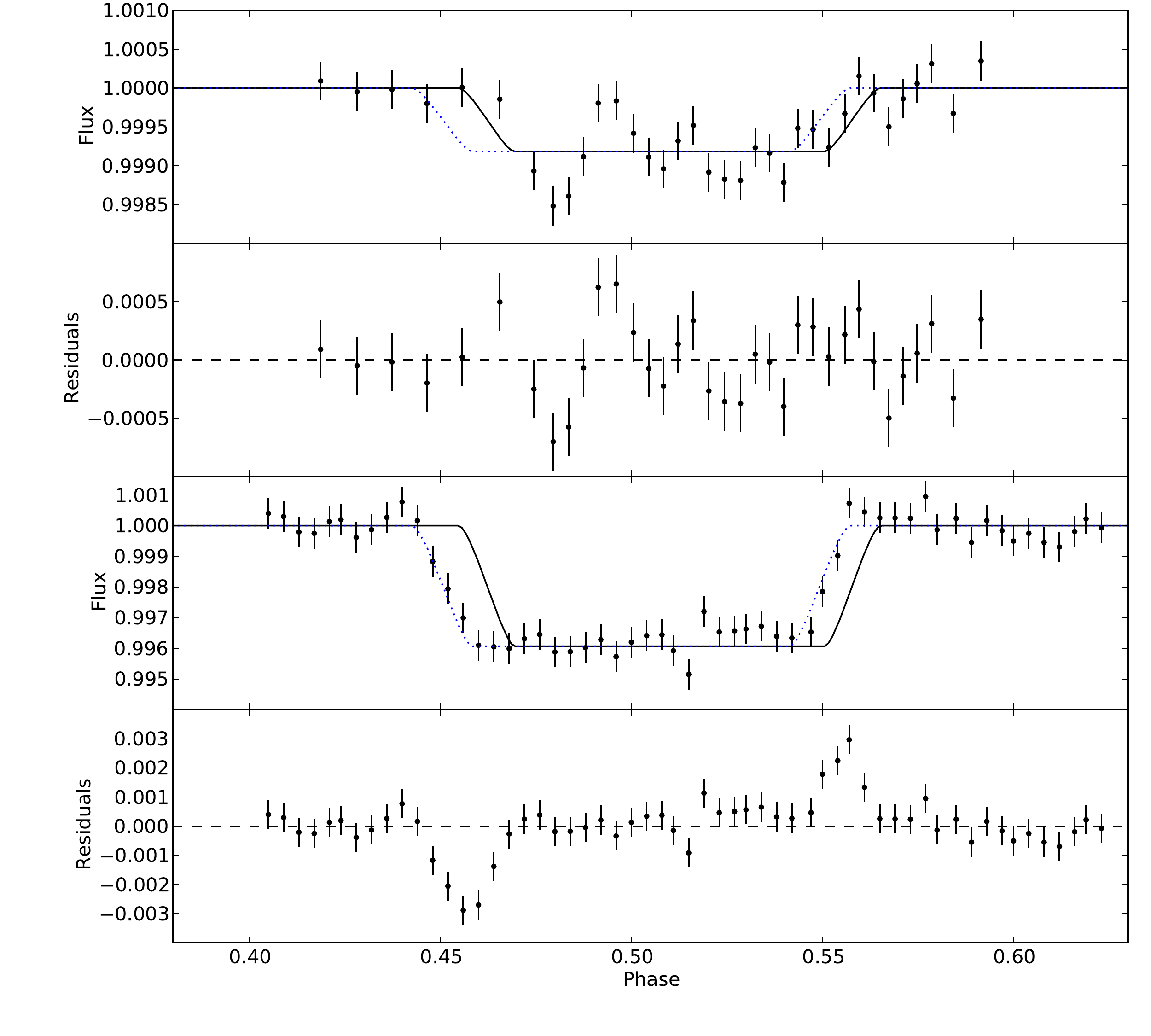}}\resizebox{8cm}{!}{\includegraphics{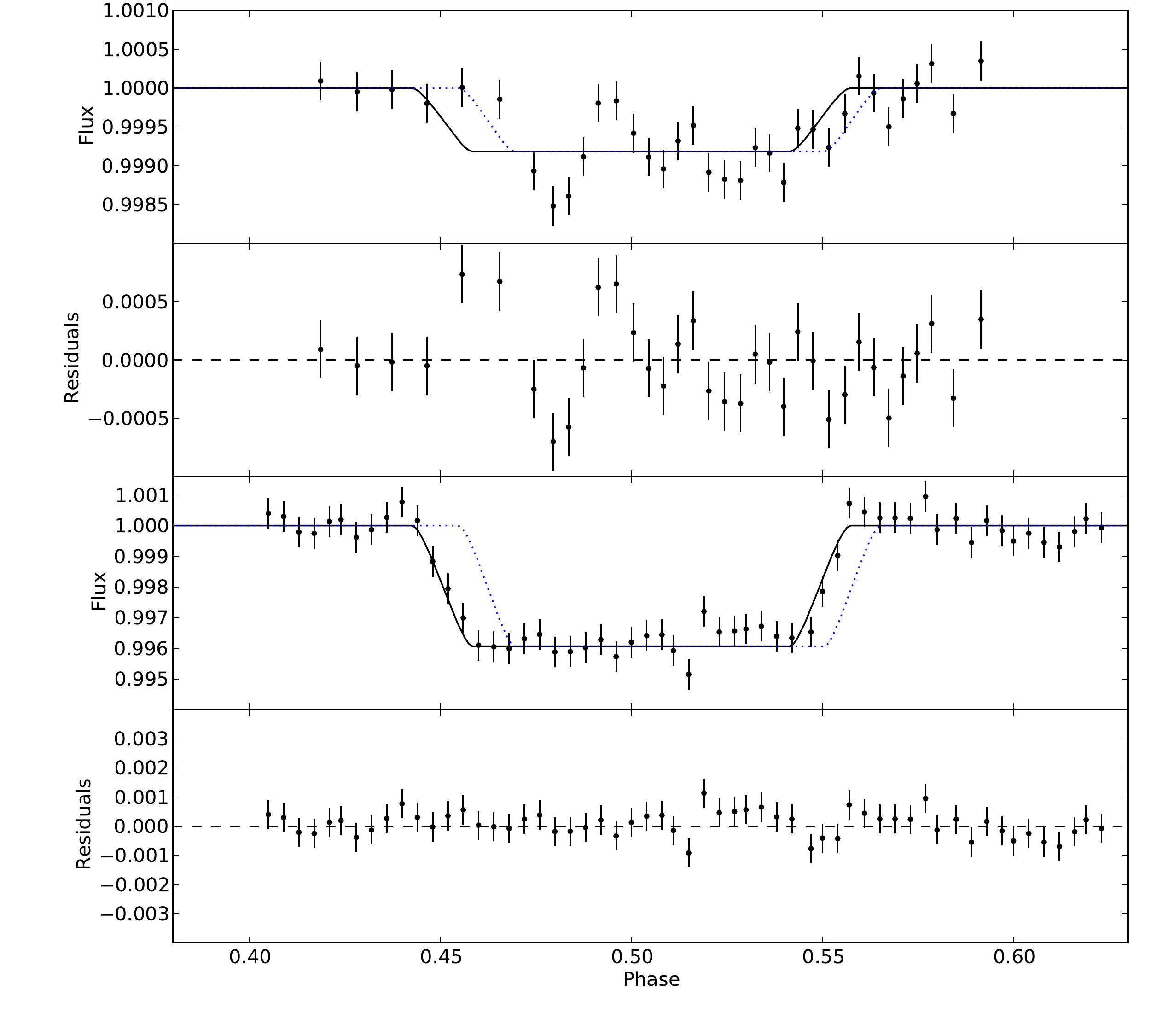}}
\caption{Eclipse flux for WASP-12. {\bf Left--} Top two panels: L09 flux and residuals. Bottom two panels: C10 flux and residuals. Solid lines represent the eclipse prediction for an eccentric orbit (L09 et al. $e=0.057$), and dotted lines represent the eclipse prediction for a circular orbit ($e=0$). Residuals are plotted for the L09 solution of $e=0.057$, which is clearly inappropriate for the C10 eclipse (bottom). {\bf Right--} Top two panels: L09 flux and residuals. Bottom two panels: C10 flux and residuals. Solid lines represent the eclipse prediction for a circular orbit ($e=0$) and  dotted lines  represent the eclipse prediction for an eccentric orbit (L09 $e=0.057$). Residuals are plotted for the circular orbit.}
\label{morales-transit}
\label{campo-transit}
\end{figure*}

It is interesting to compare the situation in radial velocity with the similar sequence of events regarding the occultation photometric data. Figure~\ref{morales-transit} (left panel) shows the eclipse data for both L09 (top two panels) and C10 (bottom two panels). The solid line represents the L09 solution, and the residuals are plotted for both datasets. The L09 solution is somewhat plausible for the L09 data but definitely not for the C10 data. The dotted line shows the a circular orbit $e=0$. Figure~\ref{campo-transit} (right panel) shows the same data, but the solid line is now a circular orbit $e=0$, and the residuals for both datasets are again plotted. It is clear that the circular solution fits the C10 dataset and remains reasonable for the L09 dataset. This could be explained if the effects of instrumental systematics had been underestimated in L09.


\subsection{The orbital eccentricity of {\small WASP}-14}
\begin{figure}
\resizebox{8cm}{!}{\includegraphics{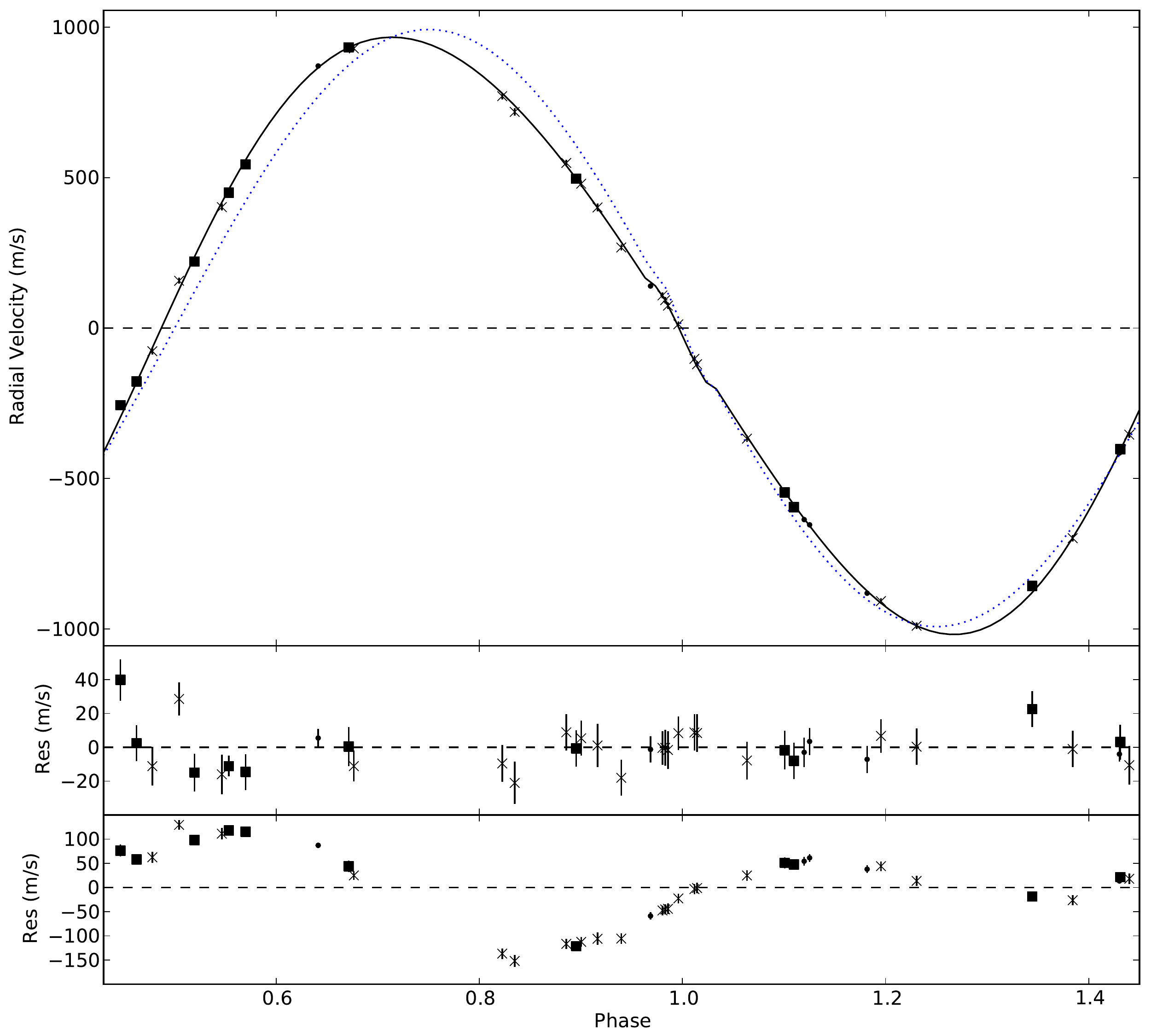}}
\caption{The new SOPHIE radial velocity data for WASP-14 are shown with squares, while the \citet{Joshi2009} data from FIES and SOPHIE are shown with points and crosses respectively. The solid line is the best solution with $e=0.088$, and the dotted line is a circular orbit. The middle panel shows residuals plotted for the best-fit orbit e=$0.088$, while the bottom panel shows residuals for a circular orbit \citep[ (using the parameters of ][ for the spectroscopic transit)]{Joshi2009}. }
\label{fig:wasp14}
\end{figure}

Figure~\ref{fig:wasp14}  shows our SOPHIE data and the FIES and SOPHIE data from \citet{Joshi2009} along with with the best-fit orbit and a circular orbit. We adopt the prior distribution on the period $P$ from photometric data by \cite{Johnson2009} and that on the mid-transit time $T_{tr}$ from \cite{Joshi2009}. We include their value of $R_{p}/R_{s}$, $a/R_{s}$ and $v\sin i$ to compute the Rositter-McLaughlin effect, and estimate the quadratic limb darkening coefficients $u_a=0.405$ and $u_b=0.337$ from \cite{Claret2004}. We work out the orbital parameters using both our radial velocity data and that published in the discovery paper (since we do not have any radial velocity data during the transit, the RM effect from that paper are not significantly modified). We include the effect of red noise in a similar manner explained above, adding $8$ \ms\ of correlated uncertainties to each single datapoint and $\sqrt{N}\times 8$ \ms\ to each point in each group of $N$ points taken on the same night. 

 Our results are shown in Table~\ref{tab:wasp14}. Our best-fit value for the orbital eccentricity is $e=0.088\pm0.003$, in good agreement with the value found by  \cite{Joshi2009}. 

We examined the possibility of a scenario similar to that of WASP-12, with correlated noise causing a spurious eccentricity detection. The lower panel of Figure~\ref{fig:wasp14} shows the residuals around the best-fit circular orbit (given by a MCMC run with $e=0$ fixed). We find that, in contrast to the case of WASP-12, the differences between observation and model assuming a circular solution are periodic and regular, which would not be the case for correlated noise. We also looked for possible evidence for a second planetary companion in the system by examining the residuals as  function of time, finding no unambiguous trend. 

 
\begin{table*}
\centering
\begin{tabular}{ l c c  }
\hline
Parameter										& Joshi et al				& This paper \\ \hline
Centre-of-mass velocity  $V_0$[\kms]  	& 						& $-$4.985$\pm$0.003 \\
Orbital eccentricity $e$ 								& 0.091$\pm$0.003		& 0.088$\pm$0.003 \\
Argument of periastron $\omega$ [$^o$]				& $-$106.6$\pm$0.7			& $-$107$\pm$1\\


Semi-amplitude K [\ms]				& 993$\pm$3				& 991$\pm$3 \\ \hline
%
\end{tabular}
\caption{System parameters for WASP-14. (Left: Joshi et al. Right: Our results.) }
\label{tab:wasp14}
\end{table*}

\section{Discussion}



Our results confirm the strong indications of C10 that all the available data for WASP-12 is compatible with a circular orbit, and that the eccentricity of the best-fit orbit to the radial velocity of H09 and subsequently the occultation data of L09  may be due to correlated noise. Not accounting for this noise in the statistical analysis could lead to an apparent $\sim 3\sigma$ significance for the rejection of the null hypothesis ($e=0$), but the new data strongly suggest that the orbit of WASP-12b is indeed circular. 

C10 suggested that, for an eccentric orbit, the difference between the eclipse phase in C10 and L09 could be due to apsidal precession. This would require that the argument of periastron had changed from $\omega=-74^{+13}_{-10}$ in Feb 2008 (H09) to $\omega\sim -90^{o}$ in Oct 2008 (C10). In terms of the projected component of the eccentricity, $e\cos\omega$, one obtains $e\cos\omega=0.014\pm{0.004}$ in Feb 2008 using the H09 orbital parameters, C10 found $e\cos\omega=0.0019\pm{0.0007}$ in Oct 2008, L09 found $e\cos\omega=0.0156\pm{0.0035}$ in Feb-Oct 2009, and our result is $e \cos\omega = 0.0037\pm0.0035$ for January 2009 to March 2010. When considered in the order that observations were made, these values do not support the hypothesis of apsidal precession. 



A circular orbit for WASP-12 removes the needs for models to explain the survival of such an eccentricity at this very short period, in face of what would have been extremely strong tidal effects. In particular, the scenario of \citet{Li2010}, using the eccentricity from H09 to infer values of mass loss and tidal dissipation for WASP-12, lose its principal empirical support. 


\begin{figure}
\resizebox{8cm}{!}{\includegraphics{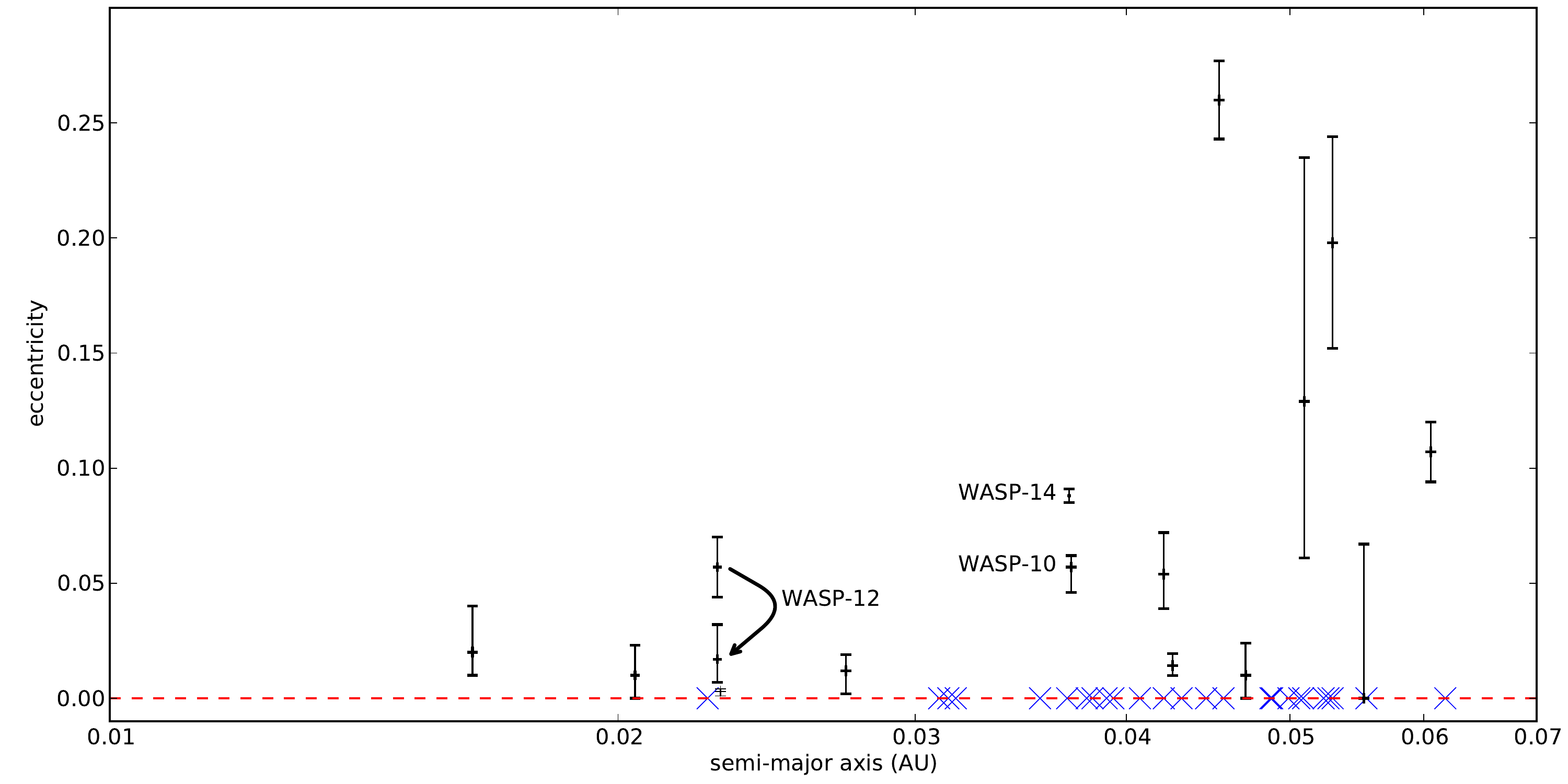}}
\caption{Plot of eccentricity $e$ against semi-major axis $a$ (log scale) for close-in transiting planets. 
Orbits with measured eccentricities are shown with a plus sign, and orbits with assumed zero eccentricities with a cross. \newline
$^*$ \citet{Christian2009}.}
\label{fig:ecc}
\end{figure}


The eccentricity of WASP-14b, by contrast, is solidly confirmed by our measurements. This illustrates the capacity of SOPHIE to measure accurate values of orbital eccentricity for transiting planets, given a sufficient number of measurements well distributed in phase and spread over different nights (the measurements for WASP-12 having most weight towards an eccentric solution were gathered during only two different nights).

\subsection*{Eccentricity distribution and tidal circularisation for hot Jupiters}

Figure~\ref{fig:ecc} shows a plot of the eccentricity of  known transiting planets with $a<0.1{\rm AU}$ against their semi-major axis $a$ (log scale). There is a well-know trend for the inner planets to have circular orbits, due to tidal orbital decay. If the WASP-12 system follows a circular orbit, the WASP-14 system establishes a new lower limit for the point at which orbital eccentricity can survive tidal evolution for a sufficiently long time to be observed in a sample not selected by age. The fact that the eccentricity of WASP-14 is clearly non-zero, but also near the lower end of the eccentricity distribution for planets out of reach of tidal evolution (broadly distributed between 0 and 1) probably indicates that this object has already undergone some tidal evolution. As a result, WASP-14 is an interesting object to constrain the tidal evolution, intensity of tidal torque and relation between tides and size, for gas giant planets.

WASP-10 is a third known transiting planets closer than 0.04 AU with a Êpossible non-circular orbit. However, the eccentricity found by Christian et al. (2009) is significant to a level similar to H09 for WASP-12, and based in a comparable number and quality of RV measurements.  The evidence for a circular orbit actually rests on two SOPHIE measurements, with residuals compared to a circular orbit that are comparable to the amplitude of the instrumental effects found in the High Efficiency mode.  It is therefore possible that this detection too is due to underestimated instrumental systematics acting on a circular orbit, in the manner described in \citet{Lucy1971}.

The fact that WASP-12b has a large size ($R_p=1.79\,{\rm R_J}$, for $a=0.0229$ and $e\sim 0$) and WASP-14b a smaller but still markedly inflated size ($R_p=1.28\,{\rm R_J}$, for $a=0.037$ AU and $e\sim 0.09$), provides a challenging test for theories attempting to account for the anomalous radii of some hot jupiters.




\subsection*{Acknowledgements}

We thank the SOPHIE Exoplanet Consortium for arranging a flexible observation schedule for our programme, and the whole OHP/SOPHIE team for support.

\begin{table}
\centering
\begin{tabular}{ l r r }
\hline
Time  & RV   & $\sigma_{\rm RV}$ \\
\ [HJD-2450000] &  [\kms] &[\kms]  \\
\hline
4849.41844 & 19.2172 & 0.0119\\
4849.42764 & 19.2028 & 0.0112\\
4849.43654 & 19.1811 & 0.0111\\
4849.44552 & 19.1807 & 0.0111\\
4849.45469 & 19.1678 & 0.0108\\
4849.46419 & 19.1685 & 0.0106\\
4849.47405 & 19.1444 & 0.0106\\
4849.48469 & 19.1353 & 0.0108\\
4849.49611 & 19.0984 & 0.0118\\
4849.50751 & 19.0857 & 0.0121\\
4849.51895 & 19.0812 & 0.0123\\
4849.53036 & 19.0532 & 0.0123\\
4849.54177 & 19.0163 & 0.0121\\
4849.55319 & 19.0076 & 0.0129\\
4849.56461 & 19.0083 & 0.0119\\
4849.57603 & 18.9966 & 0.0132\\
4889.37950 & 19.1295 & 0.0040 \\
4890.44193 & 19.0795 & 0.0047\\
4912.37616 & 19.1793 & 0.0085\\
4914.36671 & 18.9750 & 0.0044\\
4926.34294 & 18.9295 & 0.0043\\
4935.32914 & 19.2684 & 0.0118\\
4936.31657 & 19.1644 & 0.0068\\
5269.37102 & 19.3050 & 0.0038\\
5271.33688 & 19.0719 & 0.0039\\
5272.38229 & 19.0430 & 0.0039\\
5273.33965 & 18.8919 & 0.0039\\
5282.31904 & 19.1749 & 0.0045\\
5283.35164 & 19.1109 & 0.0049\\ \hline
\end{tabular}
\caption{Radial velocity measurements for WASP-12 (errors include random component only) }
\label{tab:rv_wasp12}
\end{table}

\begin{table}
\centering
\begin{tabular}{ l r r }
\hline
Time  & RV   & $\sigma_{\rm RV}$ \\
\ [HJD-2450000] &  [\kms] &[\kms]  \\
\hline
4885.65165 & -5.5847 & 0.0107\\
4886.57106 & -4.7678 & 0.0111\\
4886.68397 & -4.4453 & 0.0106\\
4888.61655 & -5.3910 & 0.0104\\
4888.68715 & -5.1660 & 0.0107\\
4889.65818 & -4.4914 & 0.0107\\
4890.66568 & -5.8442 & 0.0106\\
4911.59331 & -4.0451 & 0.0116\\
4912.55658 & -5.5238 & 0.0114\\
4913.57244 & -4.5271 & 0.0061\\
4915.57681 & -5.2325 & 0.0122\\ \hline
\end{tabular}
\caption{Radial velocity measurements for WASP-14  (errors include random component only) }
\label{tab:rv_wasp14}
\end{table}

\bibliography{husnoo}{}

\begin{thebibliography}{}

\bibitem[\protect\citeauthoryear{{Bouchy}, {Hebrard}, {Udry} et~al.,}{{Bouchy}
  et~al.}{2009}]{Bouchy2009}
{Bouchy} F.,  {Hebrard} G.,  {Udry} S.,    et~al., 2009, A\&A, 505, 853

\bibitem[\protect\citeauthoryear{{Campo}, {Harrington}, {Hardy}
  et~al.,}{{Campo} et~al.}{2010}]{Campo2010}
{Campo} C.~J.,  {Harrington} J.,  {Hardy} R.~A.,    et~al., 2010,
  ArXiv:1003.2763

\bibitem[\protect\citeauthoryear{{Christian}, {Gibson}, {Simpson}
  et~al.,}{{Christian} et~al.}{2009}]{Christian2009}
{Christian} D.~J.,  {Gibson} N.~P.,  {Simpson} E.~K.,    et~al., 2009, \mnras,
  392, 1585

\bibitem[\protect\citeauthoryear{{Claret}}{{Claret}}{2004}]{Claret2004}
{Claret} A.,  2004, A\&A, 428, 1001

\bibitem[\protect\citeauthoryear{Gim{\'e}nez}{Gim{\'e}nez}{2006}]{Gimenez2006}
Gim{\'e}nez A.,  2006, apj, 650, 408

\bibitem[\protect\citeauthoryear{{Hebb}, {Collier-Cameron}, {Loeillet}
  et~al.,}{{Hebb} et~al.}{2009}]{Hebb2009}
{Hebb} L.,  {Collier-Cameron} A.,  {Loeillet} B.,    et~al., 2009, \apj, 693,
  1920

\bibitem[\protect\citeauthoryear{{H{\'e}brard}, {Bouchy}, {Pont}
  et~al.,}{{H{\'e}brard} et~al.}{2008}]{Hebrard2008}
{H{\'e}brard} G.,  {Bouchy} F.,  {Pont} F.,    et~al., 2008, \aap, 488, 763

\bibitem[\protect\citeauthoryear{{Holman}, {Winn}, {Latham} et~al.,}{{Holman}
  et~al.}{2006}]{holman2006}
{Holman} M.~J.,  {Winn} J.~N.,  {Latham} D.~W.,    et~al., 2006, \apj, 652,
  1715

\bibitem[\protect\citeauthoryear{Johnson, Winn, Albrecht et~al.,}{Johnson
  et~al.}{2009}]{Johnson2009}
Johnson J.~A.,  Winn J.~N.,  Albrecht S.,    et~al., 2009, Publications of the
  Astronomical Society of the Pacific, 121, 1104

\bibitem[\protect\citeauthoryear{{Joshi}, {Pollacco}, {Cameron}
  et~al.,}{{Joshi} et~al.}{2009}]{Joshi2009}
{Joshi} Y.~C.,  {Pollacco} D.,  {Cameron} A.~C.,    et~al., 2009, \mnras, 392,
  1532

\bibitem[\protect\citeauthoryear{{Li}, {Miller}, {Lin} et~al.,}{{Li}
  et~al.}{2010}]{Li2010}
{Li} S.,  {Miller} N.,  {Lin} D.~N.~C.,    et~al., 2010, \nat, 463, 1054

\bibitem[\protect\citeauthoryear{{Loeillet}, {Shporer}, {Bouchy}
  et~al.,}{{Loeillet} et~al.}{2008}]{Loeillet2008}
{Loeillet} B.,  {Shporer} A.,  {Bouchy} F.,    et~al., 2008, A\&A, 481, 529

\bibitem[\protect\citeauthoryear{{Lopez-Morales}, {Coughlin}, {Sing}
  et~al.,}{{Lopez-Morales} et~al.}{2009}]{Morales2009}
{Lopez-Morales} M.,  {Coughlin} J.~L.,  {Sing} D.~K.,    et~al., 2009,
  arXiv:0912.2359

\bibitem[\protect\citeauthoryear{{Lucy} \& {Sweeney}}{{Lucy} \&
  {Sweeney}}{1971}]{Lucy1971}
{Lucy} L.~B.,  {Sweeney} M.~A.,  1971, \aj, 76, 544

\bibitem[\protect\citeauthoryear{{Mandel} \& {Agol}}{{Mandel} \&
  {Agol}}{2002}]{Mandel2002}
{Mandel} K.,  {Agol} E.,  2002, \apjl, 580, L171

\bibitem[\protect\citeauthoryear{{Perruchot}, {Kohler}, {Bouchy}
  et~al.,}{{Perruchot} et~al.}{2008}]{Perruchot2008}
{Perruchot} S.,  {Kohler} D.,  {Bouchy} F.,    et~al., 2008, in Society of
  Photo-Optical Instrumentation Engineers (SPIE) Conference Series Vol.~7014,
  {The SOPHIE spectrograph: design and technical key-points for high throughput
  and high stability}

\bibitem[\protect\citeauthoryear{{Pont} \& {Eyer}}{{Pont} \&
  {Eyer}}{2004}]{Pont2004}
{Pont} F.,  {Eyer} L.,  2004, \mnras, 351, 487

\bibitem[\protect\citeauthoryear{{Pont}, {Hebrard}, {Irwin} et~al.,}{{Pont}
  et~al.}{2009}]{Pont2009b}
{Pont} F.,  {Hebrard} G.,  {Irwin} J.~M.,    et~al., 2009, A\&A, 502, 695

\bibitem[\protect\citeauthoryear{{Pont}, {Zucker} \& {Queloz}}{{Pont}
  et~al.}{2006}]{Pont2006}
{Pont} F.,  {Zucker} S.,    {Queloz} D.,  2006, MNRAS, 373, 231

\end{thebibliography}

\end{document}